\documentclass[journal]{IEEEtran}

\IEEEoverridecommandlockouts
\usepackage{amsmath,amssymb,amsfonts}
\usepackage{cite}
\usepackage{subfigure}
\usepackage{graphicx}
\usepackage{textcomp}
\usepackage{xcolor}
\usepackage{times}
\usepackage{subfigure}         
\usepackage{amssymb,amsmath}
\usepackage{booktabs}

\usepackage{acronym}  % make an acronym

\usepackage{balance}

\usepackage{epstopdf}

\usepackage{bm}   

\usepackage{algorithm} 

\usepackage{algorithmic}

\usepackage{arydshln}

\usepackage{lipsum}
\usepackage{stfloats}
\usepackage{url}

\usepackage{setspace}
\usepackage{hyperref}
\hypersetup{
	colorlinks = true, %
	linkcolor = black, %
	urlcolor = blue,%
	citecolor = blue} %
\allowdisplaybreaks[4]

\newcommand{\mb}[1]{{  \mathbf  #1}}  %\mathbf  \bm

\definecolor{BLUE}{rgb}{0,0,1}

\acrodef{siso}[SISO]{single-input single-output}%

\acrodef{nmse}[NMSE]{normalized mean square error}%

\acrodef{ris}[RIS]{reconfigurable intelligent surface}%

\acrodef{mm}[MM]{majorization-minimization}%
\acrodef{IF}[IF]{ice-filling}%

\acrodef{rhs}[RHS]{reconfigurable holographic surface}%

\acrodef{csi}[CSI]{channel state information}%
\acrodef{awgn}[AWGN]{additive white Gaussian noise}%

\acrodef{mim}[MIM]{mutual-information-maximization}%
\acrodef{mi}[MI]{mutual information}%

\acrodef{gpr}[GPR]{Gaussian process regression}%

\acrodef{cs}[CS]{compressed sensing}%
\acrodef{ao}[AO]{alternaing optimization}%

\acrodef{bs}[BS]{base station}%
\acrodef{snr}[SNR]{signal-to-noise ratio}%
\acrodef{mmwave}[mmWave]{millimeter-wave}%
\acrodef{snr}[SNR]{signal-to-noise ratio}%
\acrodef{rf}[RF]{radio frequency}%

\acrodef{das}[DAS]{dense array system}%

\acrodef{omp}[OMP]{orthogonal matching pursuit}%
\acrodef{mp}[MP]{message passing}%

\acrodef{ml}[ML]{maximum likelihood}%

\acrodef{mse}[MSE]{mean square error }%

\acrodef{as}[AS]{antenna selection}%

\acrodef{vamp}[VAMP]{vector approximate message passing}%

\acrodef{fas}[FAS]{fluid antenna system}%

\acrodef{sinr}[SINR]{signal-to-interference-plus-noise ratio}%

\acrodef{mimo}[MIMO]{multiple-input multiple-output}%
\acrodef{miso}[MISO]{multiple-input single-output}%
\makeatletter
\def\firstletterparse#1#2&{\def\strfirstletter{#1}\def\strotherletters{#2}}

\newcommand{\MakeSmallcaps}[1]{%
	\expandafter\firstletterparse#1&
	\expandafter\MakeUppercase\strfirstletter\textsc{\strotherletters}%
}%\textrm

\makeatother

\def\BibTeX{{\rm B\kern-.05em{\sc i\kern-.025em b}\kern-.08em
		T\kern-.1667em\lower.7ex\hbox{E}\kern-.125emX}}

\begin{document}
	\title{Enhancing Channel Estimation in RIS-aided Systems via Observation Matrix Design
	}
\author{
{Zijian Zhang and Mingyao Cui
}
\vspace{-2em}
\thanks{Zijian Zhang is with the Department of Electronic Engineering, Tsinghua University, Beijing 100084, China, as well as the Beijing National Research Center for Information Science and Technology (BNRist), Beijing 100084, China (e-mail: zhangzij15@tsinghua.org.cn).}
\thanks{Mingyao Cui is with  the Department of Electrical and Electronic Engineering, The University of Hong Kong, Hong Kong (e-mail: mycui@eee.hku.hk). {\it(Corresponding author: Mingyao Cui.)}}
}	
%\markboth{Journal of \LaTeX\ Class Files,~Vol.~18, No.~9, September~2020}%
%{How to Use the IEEEtran \LaTeX \ Templates}
%\markboth{ IEEE Transactions on Communications}{ Zhang {\it et al.}: 
%\paperTitleMarkboth}
\maketitle
%\tableofcontents
%Furthermore, the observation matrix design is extended to the phase-only controllable case, which can be addressed by a majorization-minimization (MM)-based method.
\begin{abstract}
Reconfigurable intelligent surfaces (RISs) have emerged as a promising technology for enhancing wireless communications through dense antenna arrays. Accurate channel estimation is critical to unlocking their full performance potential. To enhance RIS channel estimators, this paper proposes a novel observation matrix design scheme. Bayesian optimization framework is adopted to generate observation matrices that maximize the mutual information between received pilot signals and RIS channels. To solve the formulated problem efficiently, we develop an alternating Riemannian manifold optimization (ARMO) algorithm to alternately update the receiver combiners and RIS phase-shift matrices. An adaptive kernel training strategy is further introduced to iteratively refine the channel covariance matrix without requiring additional pilot resources. Simulation results demonstrate that the proposed ARMO-enhanced estimator achieves substantial gains in estimation accuracy over state-of-the-art methods.
\end{abstract}
%\vspace{-2em}

\section{Introduction}
Dense array systems (DASs) integrate massive sub-wavelength antennas within a compact aperture to enhance communication performance. Prominent examples include holographic multi-input multi-output (MIMO)~\cite{HuangChongWen'20'WCOM}, continuous-aperture MIMO~\cite{Zijian'23'JSAC}, super-directive antenna arrays~\cite{kanbaz2024}, and fluid antenna systems~\cite{wong2021fluid}. As a representative DAS, reconfigurable intelligent surfaces (RISs) comprise a large number of passive reflecting elements, enabling dynamic manipulation of the wireless environment within a limited aperture. Accurate channel estimation is crucial to support this manipulation~\cite{zhang2023reconfigurable}. Since RISs typically operate as passive reflectors, channel estimation must recover the high-dimensional transmitter-RIS-receiver cascaded channels. This high dimensionality leads to prohibitive pilot overhead for channel estimation~\cite{zhang2023reconfigurable}. 

{\color{black}
To reduce this overhead, numerous channel estimation methods have been proposed that exploit inherent channel structures, such as angular-domain sparsity~\cite{wang2020compressed}, quasi-static channel conditions~\cite{Huchen'21}, or common scatterers in multi-user channels~\cite{wei2021channel}. 
Particularly, to diversify the pilot symbols during channel estimation, existing schemes often employ random phase-shift codebooks \cite{Huchen'21,wei2021channel} or discrete Fourier transform (DFT) codebooks \cite{Nadeem'20} at the receivers and/or RISs to build pilot observation matrices. 
While feasible, these matrices cannot fully exploit the underlying channel properties of RIS channels, resulting in a suboptimal performance compared to the optimal estimators.
	% Despite the feasibility, these settings actually do not fully utilize the channel property of RIS channels, leading to a performance gap between existing schemes and optimal estimators. 
	% To reduce pilot overhead, many RIS channel estimators have been proposed, which usually rely on some channel structures, such as the angular-domain sparsity \cite{wang2020compressed}, the quasi-static channels \cite{Huchen'21}, common scatterers of multi-user channels \cite{wei2021channel}, etc. Particularly, to diversify the pilot symbols during the training period, existing channel estimators often assume that receivers and/or RISs employ random phase-shift codebooks \cite{Huchen'21,wei2021channel} or discrete Fourier transform (DFT) codebooks \cite{Nadeem'20} to build observation matrices. Despite the feasibility, these settings actually do not fully utilize the channel property of RIS channels, leading to a performance gap between existing schemes and optimal estimators.
}

{\color{black} 
Specifically, as a typical DAS, RIS exhibits highly correlated channels due to strong spatial correlations and electromagnetic mutual coupling.
Exploiting these correlations presents a promising pathway to reduce pilot overhead without compromising estimation accuracy, thereby enhancing existing RIS channel estimators. 
Such an estimation-enhancing principle has been adopted in other DASs, including densifying MIMO \cite{cui2024nearoptimal} and FASs \cite{zhang2023successive}. 
For instance, the ``ice-filling'' technique proposed in~\cite{cui2024nearoptimal} demonstrates that aligning the observation matrix with the dominant eigendirections of correlated channels can substantially improve estimation accuracy. In \cite{zhang2023successive}, by optimizing the antenna locations for pilot reception, a Bayesian channel estimator is proposed for FASs. However, such methods are designed for point-to-point transceiver links, which are usually not applicable to RIS-aided systems.
}

To bridge this gap, this paper proposes a novel observation matrix design scheme to enhance channel estimation in RIS-aided systems. 
We first formulate a Bayesian observation matrix design framework. It involves the joint design of the RIS phase-shift matrices and the receiver combiners to greedily maximize the \ac{mi} between the RIS channels and the received pilots. 
To solve the formulated problem efficiently, we develop an alternating Riemannian manifold optimization (ARMO) scheme that alternatively optimizes observation components, which utilizes the information of channel covariance matrix, i.e., kernel. 
Finally, to facilitate kernel acquisition, an adaptive kernel training scheme is proposed to refine the channel estimates and update the kernel iteratively.
Simulation results demonstrate that our proposed design can achieve significant gains in estimation accuracy over state-of-the-art methods.

%this scheme exploits the strong correlation of RIS-aided channels to jointly design the RIS phase-shift matrix and the receiver combiner. The main contributions include: (i) We propose a generic joint observation-matrix design framework for RIS-aided systems, explicitly enforcing the constant-modulus (ii) We develop an ARMO algorithm to efficiently solve the resulting mutual-information maximization problem by alternating on the Riemannian manifolds of feasible solutions. (iii) We introduce a coupled channel/kernel refinement procedure that iteratively improves the Gaussian-process prior and the channel estimates. 

% Importantly, the proposed design is estimator-agnostic: it can enhance any channel estimator (e.g., MMSE) by providing more informative pilot observations.

% Our method follows a GPR-based mutual-information maximization criterion: we formulate a Bayesian channel estimation model and design the RIS phase shifts and combiner coefficients to maximize the information about the channel gained per pilot. To solve this challenging non-convex design problem, we develop an Alternating Riemannian Manifold Optimization (ARMO) algorithm that alternately updates the RIS and combiner matrices on their unit-modulus manifolds. We further introduce a coupled iterative scheme that refines the channel estimates and updates the kernel (covariance) model for improved accuracy.

The remainder of this paper is organized as follows. In Section II, the system model is introduced, and the observation matrix design problem is formulated. In Section III, the proposed ARMO scheme and the kernel training scheme are provided. In Section IV, simulation results are carried out to verify the effectiveness of the proposed schemes. Finally, conclusions are drawn in Section V.

\textit{Notation:} ${[\cdot]^{T}}$, ${[\cdot]^{H}}$, ${[\cdot]^{*}}$, ${[\cdot]^{\dagger}}$, and ${[\cdot]^{-1}}$ represent the transpose, conjugate-transpose, complex conjugate, pseudo-inverse, and inverse, respectively. The operator $\|\cdot\|$ stands for the $l_2$-norm. For a matrix ${\bf Z}$, ${\bf Z}(:,j)$ indicates its $j$-th column. ${\rm Tr}(\cdot)$ specifies the trace, and ${\rm diag}(\cdot)$ constructs a diagonal matrix. The expectation operator is given by ${\mathsf E}(\cdot)$, and $\Re({\cdot})$ extracts the real part. A complex Gaussian distribution with mean ${\bm \mu}$ and covariance matrix ${\bf \Sigma}$ is written as $\mathcal{C}\mathcal{N}({\bm \mu}, {\bf \Sigma})$. $\mathbf{I}_{L}$ designates the $L\times L$ identity matrix, $\mathbf{1}_{L}$ an all-ones vector or matrix of dimension $L$, and $\mathbf{0}_{L}$ a zero vector of dimension $L$.

\section{System Model and Problem Formulation}\label{sec:model}
In this section, the system model is first introduced in Subsection \ref{subsec:model}, Then, the problem of observation matrix design is formulated in Subsection \ref{subsec:problem}.

\subsection{System Model}\label{subsec:model}
This paper considers the uplink channel estimation of an RIS-aided DAS, which is composed of an $M$-antenna \ac{bs} with one \ac{rf} chain, an $N$-element RIS, and a single-antenna user \cite{cui2024nearoptimal}. The antenna spacing of \ac{bs} and the element spacing of RIS are both on the order of sub-wavelength. The user-RIS channel and the RIS-\ac{bs} channel are defined as ${\bf g}\in{\mathbb C}^{N}$ and ${\bf F}^H\in{\mathbb C}^{M \times N}$, respectively. Let $Q$ denote the number of pilots transmitted by the user in a coherence-time frame. The $q$-th received pilot at the \ac{bs} can be modeled as
\begin{align}\label{eqn:y_p}
{y_q} = {\bf{w}}_q^H{{\bf{F}}^H}{{\bf{\Theta }}_q}{\bf{g}}s_q + {n_q},
\end{align}
where vector ${\bf{w}}_q\in{\mathbb C}^M$ is the combiner at the \ac{bs}; diagonal matrix ${\bf{\Theta }}_q:={\rm diag}(e^{j\theta_1},\cdots,e^{j\theta_N})\in{\mathbb C}^{N\times N}$ is the phase shift matrix at the RIS; $s_q$ 
is the pilot symbol; and ${n}_q \sim \mathcal{C}\!\mathcal{N}\left({\bf 0}, \sigma^2 \right)$ is the \ac{awgn}. Owing to the fully analog architecture, the combiner and phase shift matrix should satisfy the constant-modulus constraints: $|{\bf{w}}_q(m)|=1/\sqrt{M}$ for all $m\in\{1,\cdots, M\}$ and $|{\bf{\Theta}}_q(n,n)|=1$ for all $n\in\{1,\cdots, N\}$, respectively. 

Without loss of generality, we assume that $s_q = 1$ for all $q\in\{1,\cdots,Q\}$. It can be easily proved that, by using some matrix techniques, the original model (\ref{eqn:y_p}) can be rewritten as 
\begin{align}\label{eqn:y_p_revised}
{y_q} = \left( {{\bf{w}}_q^H \otimes {\bm{\theta }}_q^T} \right){\bf{h}} + {n_q} = {\bf x}_q^H{\bf{h}} + {n_q},
\end{align}
where ${\bm{\theta }}_q:=[e^{j\theta_1},\cdots,e^{j\theta_N}]^T$ is defined as the phase shift vector; ${\bf x}_q:={{\bf{w}}_q \otimes {\bm{\theta }}_q^*}\in {\mathbb C}^{MN}$ is the observation vector at timeslot $q$; and $\bf h$ is the vectorized equivalent user-RIS-\ac{bs} channel, defined as
\begin{align}\label{eqn:h_equiv}
{\bf{h}} = {\left[ {{{\bf{g}}^H}{\rm{diag}}\left( {\bf{F}}\left( {:,1} \right) \right), \cdots ,{{\bf{g}}^H}{\rm{diag}}\left( {\bf{F}}\left( {:,M} \right) \right)} \right]^H},
\end{align}
which is exactly the cascaded channel to be estimated for subsequent beamforming design. Since $Q$ timeslots are utilized for channel estimation, the overall received pilots can be expressed as 
\begin{equation}\label{eqn:y}
	{\bf y} = {\bf X}^{H}{\bf h} + {\bf n},
\end{equation}
where ${\bf y} := [{y}_1^T,\cdots,{y}_Q^T]^T$,  ${\bf n} := 
[ {\bf n}_1^T,\cdots, {\bf n}_Q^T ]^T$, and ${\bf{X}} = 
\left[{\bf{x}}_1,\cdots, {\bf{x}}_Q\right]$. This work aims to jointly design combiners $\{\mb{w}_q\}_{q=1}^Q$ and precoders $\{{\bm \theta}_q\}_{q=1}^Q$ so as to better recover channel $\bf h$ from the received pilots $\bf{y}$.

\subsection{Problem Formulation}\label{subsec:problem}
Since the antenna spacing of the \ac{bs} and the element spacing of the RIS are small, the cascaded user-RIS-\ac{bs} channel is strongly correlated due to the high spatial correlations and the electromagnetic mutual coupling among antennas and elements. Let ${\bf{\Sigma }}_{\bf{g}}:={\mathsf E}({\bf{g}}{\bf{g}}^H)\in{\mathbb C}^{N\times N}$ denote the covariance of channel $\bf g$ and ${\bf{\Sigma }}_{\bf{F}}:={\mathsf E}({\rm vec}({\bf{F}}){{\rm vec}({\bf{F}})}^H)\in{\mathbb C}^{MN\times MN}$ denote the covariance of channel $\bf F$, respectively. Assuming the mean of $\bf h$ is zero, the covariance of the equivalent channel $\bf h$ can be derived from (\ref{eqn:h_equiv}) as
$
{{\bf{\Sigma }}_{\bf{h}}} := {\mathsf E}\left( {{\bf{h}}{{\bf{h}}^H}} 
\right) = {\bf{\Sigma }}_{\bf{F}}^* \odot \left( {{{\bf{1}}_{M,M}} \otimes {\bf{\Sigma }}_{\bf{g}}} \right),
$
which is known as the {\it kernel} of channel. 
The high spatial correlation and the mutual coupling indicate that the kernel $\bm{\Sigma_{\bf h}}$ is structured and underdetermined, which can provide prior knowledge for achieving high-accuracy channel estimation \cite{Sungwoo'18, Karthik'18}. To realize this potential, we adopt the idea of \ac{gpr} to design the observation matrix $\bf X$. Specifically, the channel is modeled as being sampled from the Gaussian process ${\cal CN}\left({\bf 0}_{MN}, {\bm \Sigma}_{\bf h}\right)$. The joint probability distribution of $\mb{h}$ and $\mb{y}$ then satisfies
\begin{equation}
	\begin{aligned}
		\left[ \begin{array}{l}
			{\bf{h}}\\
			{{\bf{y}}}
		\end{array} \right] \sim \mathcal{CN} \left(
		\left[ \begin{array}{c}
			{\bf 0}_{MN} \\
			{\bf 0}_{Q}
		\end{array} \right], 
		\left[ 
		{\begin{array}{*{20}{c}}
				{{{\bf{\Sigma }}_{\bf{h}}}}&{{\bf{\Sigma 
					}}_{\bf{h}}}{{\bf{X}}}\\{{\bf{X}}^H}
				{{\bf{\Sigma 
					}}_{\bf{h}}}& {\bf X}^{H}{\bf{\Sigma 
				}}_{\bf{h}}{\bf X} + \sigma^2{\bf I}_Q
		\end{array}} \right]\right).
	\end{aligned}
\end{equation}
Thereby, 
the posterior mean and the posterior covariance 
of $\bf h$ are expressed as
\begin{equation}\label{eq:postmean}
	\hat{\bf h} := {{\bm{\mu }}}_{\mb{h}|\mb{y}} = {{\bf{\Sigma 
		}}_{\bf{h}}}{{\bf{X}}}
	{\left( {\bf X}^{H}{\bf{\Sigma 
		}}_{\bf{h}}{\bf X} + \sigma^2{\bf I}_Q \right)^{ - 
			1}} {{\bf{y}}},
\end{equation}
\begin{equation}
	{{\bf{\Sigma }}}_{\mb{h}|\mb{y}} = {{\bf{\Sigma 
		}}_{\bf{h}}} -
	{{\bf{\Sigma 
		}}_{\bf{h}}}{{\bf{X}}}
	{\left( {\bf X}^{H}{\bf{\Sigma 
		}}_{\bf{h}}{\bf X} + \sigma^2{\bf I}_Q \right)^{ - 
			1}}{ 
		{{{\bf{X}}^H}} }
	{{\bf{\Sigma }}_{\bf{h}}},
\end{equation}
which yield the optimal channel estimator $\hat{\bf h} := {{\bm{\mu
}}}_{\mb{h}|\mb{y}}$ and the corresponding estimation error ${{\bf{\Sigma }}}_{\mb{h}|\mb{y}}$. 

Notably, the posterior covariance ${{\bf{\Sigma }}}_{\mb{h}|\mb{y}}$ depends strongly on the observation matrix $\mb{X}$. Hence, carefully designed combiners and precoders $\{\mb{w}_q\}_{q=1}^Q$ and $\{\bm{\theta}_q\}_{q=1}^Q$, i.e., $\{\mb{x}_q\}_{q=1}^Q$, can significantly reduce the channel estimation error. Motivated by this, \ac{gpr} aims to generate observation matrices that extract as much information about $\mb{h}$ as possible from the received signal $\mb{y}$. Following this principle, our objective is to find $\{\mb{w}_q\}_{q=1}^Q$ and $\{\bm{\theta}_q\}_{q=1}^Q$ that maximize the \ac{mi} between $\mb{y}$ and $\mb{h}$, which is formulated as
\begin{align}\label{eq:Entropy}
    \notag
	 &\max_{\{\mb{w}_q\}_{q=1}^Q, \{\bm{\theta}_q\}_{q=1}^Q}~I(\mb{y}; \mb{h}) 
	= 
	\log_2\det\left( \mb{I}_{Q} +  \frac{1}{\sigma^2}\mb{X}^H{\bf 
		\Sigma_h} \mb{X} \right) \\
    \notag
    &~~~~~~~~{\rm s.t.}~~~~~~|\mb{w}_q|=\frac{1}{\sqrt{M}}{\bf 1}_M,~~\forall q\in\{1,\cdots,Q\},\\ & ~~~~~~~~~~~~~~~~~~|\bm{\theta}_q|={\bf 1}_N,~~\forall q\in\{1,\cdots,Q\},
\end{align}
which resembles a classical water-filling problem in MIMO precoding. Note that, solving this problem faces two challenges. First, the coupled combiners $\{\mb{w}_q\}_{q=1}^Q$ and phase-shift vectors $\{\bm{\theta}_q\}_{q=1}^Q$ in observation matrix leads to a non-convex objective function $I(\mb{y}; \mb{h})$. Second, the constant-modulus constraints make it difficult to adopt conventional numerical algorithms. Besides, at the early stage of algorithmic implementations in practice, the knowledge of kernel ${\bf \Sigma_h}$ should be determined.

\section{Proposed Observation Matrix Design}\label{sec:2DIF}
To solve the above problem, in this section, we first focus on the joint design of $\{\mb{w}_q\}_{q=1}^Q$ and $\{\bm{\theta}_q\}_{q=1}^Q$ for a given kernel ${\bf \Sigma_h}$ in Subsection \ref{subsec:PCD_Greedy}. Then, the acquisition of ${\bf \Sigma_h}$ is addressed in Subsection \ref{subsec:Kernel_Acq}.

\subsection{Observation Matrix Design Using Greedy Method}\label{subsec:PCD_Greedy}
Due to the above challenges, obtaining the globally optimal solution to problem (\ref{eq:Entropy}) is intractable. To address this issue, we propose a greedy method that designs $\bf X$ in a column-by-column manner, i.e., generating $\{\mb{w}_q\}_{q=1}^Q$ and $\{\bm{\theta}_q\}_{q=1}^Q$ sequentially on a pilot-by-pilot basis. Specifically, we define ${{\bf{X}}_t} = [{\bf{x}}_1, 
{\bf{x}}_2, \cdots, {\bf{x}}_t]$ as the overall observation matrix for timeslots from $1$ to $t$. Let ${\bf{y}}_{t} 
= {\bf{X}}_{t}^H\mb{h} + {\bf{n}}_{t}$ denote the corresponding 
received signal, wherein ${{\bf{y}}_t} = [{{y}}_1, 
{{y}}_2, \cdots, {{y}}_t]^T$ and ${\bf{n}}_{t} := \left[ n_1,\cdots, n_t \right]^T$. 
Given the current combiners $\{\mb{w}_q\}_{q=1}^t$ and phase-shift vectors $\{\bm{\theta}_q\}_{q=1}^t$ in the first $t$ timeslots, our greedy strategy aims to find the 
combiner $\mb{w}_{t+1}$ and the phase-shift matrix $\bm{\theta}_{t+1}$ in the next timeslot, such that the \ac{mi} increment $\Delta I_{t+1} := I({\bf{y}}_{t+1};\mb{h}) - 
I({\bf{y}}_t; \mb{h})$ from timeslot $t$ to $t+1$ is maximized. By calculating \ac{mi} increment $\Delta I_{t+1}$, the $(t+1)$-th subproblem can be formulated as
\begin{align}\label{eq:MI_incre_Max}
\notag
&\mathop {\max }\limits_{\mb{w}_{t+1},\bm{\theta}_{t+1}}
\Delta I_{t+1} = \log_2\left(1+{\bf x}_{t+1}^H{\bf \Sigma}_t{\bf x}_{t+1}\right) \\
&~~~~{\rm s.t.}~~~~|\mb{w}_{t+1}|=\frac{1}{\sqrt{M}}{\bf 1}_M,~|\bm{\theta}_{t+1}|={\bf 1}_N,
\end{align}
where ${\bf x}_{t+1}:={{\bf{w}}_{t+1} \otimes {\bm{\theta }}_{t+1}^*}\in {\mathbb C}^{MN}$ is the observation vector at timeslot $q$. In particular, ${\bf \Sigma}_t$ is the posterior kernel of channel $\bf h$, which can be updated by
\begin{align}
\notag
{{\bf{\Sigma }}}_{t} & = {{\bf{\Sigma 
		}}_{\bf{h}}} -
	{{\bf{\Sigma 
		}}_{\bf{h}}}{{\bf{X}}_t}
	{\left( {\bf X}_t^{H}{\bf{\Sigma 
		}}_{\bf{h}}{\bf X}_t + \sigma^2{\bf I}_t \right)^{ - 
			1}}{ 
		{{{\bf{X}}_t^H}} }
	{{\bf{\Sigma }}_{\bf{h}}} \\
    & \overset{(a)}{=} {{\bf{\Sigma }}}_{t-1} - \frac{{{\bf{\Sigma }}}_{t-1}{\bf x}_t{\bf x}_t^H{{\bf{\Sigma }}}_{t-1}}{{\bf x}_t^H{{\bf{\Sigma }}}_{t-1}{\bf x}_t+\sigma^2},
\end{align}
where $(a)$ holds according to \cite[Appendix A]{cui2024nearoptimal}. However, due to the coupled combiner $\mb{w}_{t+1}$ and the phase-shift vector $\bm{\theta}_{t+1}$ as well as the constant-modulus constraint, finding the optimal solution to problem (\ref{eq:MI_incre_Max}) is challenging. As a suboptimal strategy, an ARMO scheme is proposed to jointly design $\mb{w}_{t+1}$ and $\bm{\theta}_{t+1}$, which is summarized in {\bf Algorithm 1}. The algorithmic details are explained as follows.

\begin{algorithm}[!t]
	%\setstretch{1.25}
	\caption{ARMO scheme for observation matrix design} 
	\begin{algorithmic}[1]\label{alg:ARMO}
		\REQUIRE  %算法的输入参数：Input
		Number of pilots $Q$, kernel ${\bm \Sigma}_{\bf h}$.
		% \vspace{3pt}
		\STATE Initialize ${\bf{\Sigma }}_{0} = {\bf{\Sigma }}_{\bf{h}}$
		\FOR{$t = 0, \cdots, Q-1$}
        \STATE Randomly generate $\mb{w}_{t+1}$ and $\bm{\theta}_{t+1}$
        \WHILE{No convergence of \ac{mi} increment $\Delta I_{t+1}$}
        \STATE Update iteration parameters: $K \leftarrow M$, ${\bf{U}} \leftarrow  ( {{{\bf{I}}_M} \otimes {\bm{\theta }}_{t + 1}^T} ) {{\bf{\Sigma }}_t} ( {{{\bf{I}}_M} \otimes {\bm{\theta }}_{t + 1}^*})$, $\rho \leftarrow 1/\sqrt{M}$, $\alpha \leftarrow {K}{\lambda _{\max }}( {\bf{U}} )/4$, $\beta  \leftarrow 1/( {{\lambda _{\max }}( {\bf{U}} ) + 2\alpha } )$
        \WHILE{No convergence of $f(\mb{w}_{t+1})$}
        \STATE Update Euclidean gradient $\nabla f({{\bf{w}}_{t + 1}})$ via (\ref{eq:Euclidean_gradient})
        \STATE Update Riemannian gradient ${{\bf{d}}_S}\left( {{{\bf{w}}_{t + 1}}} \right)$ via (\ref{eq:Riemannian_gradient})
        \STATE ${{\bf{w}}_{t + 1}} \leftarrow \exp \left( {j\angle \left( { {{\bf{w}}_{t + 1}} + \beta {{\bf{d}}_S}\left( {{{\bf{w}}_{t + 1}}} \right)} \right)} \right)/\sqrt M $ 
        \ENDWHILE
        \STATE Update iteration parameters: $K \leftarrow N$, ${\bf{U}} \leftarrow  \sum\nolimits_{m = 1}^M \sum\nolimits_{m' = 1}^M {w_{t + 1,m}}w_{t + 1,m'}^*{\bf{\Sigma }}_{t,m
,m'}^*  $, $\rho \leftarrow 1$, $\alpha \leftarrow {K}{\lambda _{\max }}( {\bf{U}} )/4$, $\beta  \leftarrow 1/( {{\lambda _{\max }}( {\bf{U}} ) + 2\alpha } )$
        \WHILE{No convergence of $f(\bm{\theta}_{t+1})$}
        \STATE Update Euclidean gradient $\nabla f(\bm{\theta}_{t+1})$ via (\ref{eq:Euclidean_gradient})
        \STATE Update Riemannian gradient ${{\bf{d}}_S}\left( \bm{\theta}_{t+1}\right)$ via (\ref{eq:Riemannian_gradient})
        \STATE $ {{\bm{\theta }}_{t + 1}} \leftarrow \exp \left( {j\angle \left( {{{\bm{\theta }}_{t + 1}} + \beta {{\bf{d}}_S}\left( {{{\bm{\theta }}_{t + 1}}} \right)} \right)} \right)$
        \ENDWHILE
        \ENDWHILE
        \STATE Calculate observation vector: ${\bf x}_{t+1}={{\bf{w}}_{t+1} \otimes {\bm{\theta }}_{t+1}^*}$
        \STATE Calculate kernel: ${{\bf{\Sigma }}}_{t+1} = {{\bf{\Sigma }}}_{t} - \frac{{{\bf{\Sigma }}}_{t}{\bf x}_{t+1}{\bf x}_{t+1}^H{{\bf{\Sigma }}}_{t}}{{\bf x}_{t+1}^H{{\bf{\Sigma }}}_{t}{\bf x}_{t+1}+\sigma^2}$
		\ENDFOR
		\ENSURE %算法的输出：Output
		Designed combiners $\{\mb{w}_q\}_{q=1}^Q$, phase-shift vectors $\{\bm{\theta}_q\}_{q=1}^Q$, and observation matrix $\bf X$.
	\end{algorithmic}
\end{algorithm}

To achieve alternating optimization, we first reformulate the original problem (\ref{eq:MI_incre_Max}) as the following two subproblems:

{\it 1) Fix $\bm{\theta}_{t+1}$ and Optimize $\mb{w}_{t+1}$:} While fixing the phase-shift vector $\bm{\theta}_{t+1}$ and removing the unrelated parts, the original subproblem (\ref{eq:MI_incre_Max}) can be reformulated as
\begin{align}\label{eq:w_optimize}
\notag
&\mathop {\max }\limits_{\mb{w}_{t+1}}~~~~
{\bf{w}}_{t + 1}^H\left( {{{\bf{I}}_M} \otimes {\bm{\theta }}_{t + 1}^T} \right){{\bf{\Sigma }}_t}\left( {{{\bf{I}}_M} \otimes {\bm{\theta }}_{t + 1}^*} \right){{\bf{w}}_{t + 1}} \\
&~~~{\rm s.t.}~~~~~|\mb{w}_{t+1}|=\frac{1}{\sqrt{M}}{\bf 1}_M,
\end{align}
where the reformulation is achieved by utilizing equality ${{\bf{w}}_{t + 1}} \otimes {\bm{\theta }}_{t + 1}^* = \left( {{{\bf{I}}_M} \otimes {\bm{\theta }}_{t + 1}^*} \right){{\bf{w}}_{t + 1}}$.

{\it 2) Fix $\mb{w}_{t+1}$ and Optimize $\bm{\theta}_{t+1}$:} Define the ${{{\bf{\Sigma }}_{t,m,m'}}}\in{\mathbb C}^{N\times N}$ as the $(m,m')$-th block of ${{\bf{\Sigma }}_t}$. Then, the posterior kernel can be rewritten as 
\begin{align}
{{\bf{\Sigma }}_t} = \left[ {\begin{array}{*{20}{c}}
{{{\bf{\Sigma }}_{t,1,1}}}& \cdots &{{{\bf{\Sigma }}_{t,1,M}}}\\
 \vdots & \ddots & \vdots \\
{{{\bf{\Sigma }}_{t,M,1}}}& \cdots &{{{\bf{\Sigma }}_{t,M,M}}}
\end{array}} \right].
\end{align}
By fixing $\mb{w}_{t+1}$ and removing the irrelevant components, subproblem (\ref{eq:MI_incre_Max}) can be rewritten as
\begin{align}\label{eq:theta_optimize}
\notag
&\mathop {\max }\limits_{\bm{\theta}_{t+1}}
~~~{\bm{\theta }}_{t + 1}^H\left( {\sum\limits_{m = 1}^M {\sum\limits_{m' = 1}^M {{w_{t + 1,m}}w_{t + 1,m'}^*{\bf{\Sigma }}_{t,m
,m'}^*} } } \right){{\bm{\theta }}_{t + 1}} \\
&~~~{\rm s.t.}~~~~~|\bm{\theta}_{t+1}|={\bf 1}_N,
\end{align}
where $w_{t + 1,m}$ is the $m$-th entry of ${\bf w}_{t+1}$.

Observing (\ref{eq:w_optimize}) and (\ref{eq:theta_optimize}), one note that these subproblems share the same form of modules-constant quadratic programming. Thus, these two subproblems can be generalized as:
\begin{align}\label{eq:v_optimize}
\notag
&\mathop {\max }\limits_{\bf{v}}
~~~{\bf{v}}^H {\bf U} {\bf v} \\
&~~{\rm s.t.}~~~|{\bf{v}}|=\rho{\bf 1}_K,
\end{align}
where $\bf{v}$, $\bf{U}$, $\rho$, and $K$ can be set to the corresponding variables in (\ref{eq:w_optimize}) and (\ref{eq:theta_optimize}), respectively. 
Given the spherical geometric structure of the constraint set, i.e., ${\cal S}:=\{{\bf{v}}\in{\mathbb C}^K : |{\bf{v}}|=\rho{\bf 1}_K\}$, Riemannian manifold optimization perfectly matches to solve problem (\ref{eq:v_optimize}). Specifically, utilizing the relation ${\bf v}^H{\bf v}= K\rho^2$, the objective function in (\ref{eq:v_optimize}) can be equivalently replaced by 
\begin{align}
f({\bf v})={\bf{v}}^H ({\bf U}+\alpha{\bf I}_K) {\bf v},
\end{align}
where $\alpha>0$ is a positive constant to ensure the algorithmic convergence. Given $f({\bf v})$, the Euclidean gradient and the Riemannian gradient over the tangent space of ${\cal S}$ can be respectively written as:
\begin{align}
\label{eq:Euclidean_gradient}
\nabla f({\bf{v}}) & = ({\bf{U}} + \alpha {{\bf{I}}_K}){\bf{v}}, \\
\label{eq:Riemannian_gradient}
{\bf{d}}_{\cal S}\left( {\bf{v}} \right) & = {\nabla }f({\bf{v}}) - \frac{{{\bf{v}} \odot \Re \left( {{{\bf{v}}^*} \odot {\nabla }f({\bf{v}})} \right)}}{{{\rho ^2}}}.
\end{align}
In each iteration, the update over the tangent space  can be written as
\begin{align}\label{eq:V_update}
{{\bf{v}}} \leftarrow \rho \exp \left( {j\angle \left( {{{\bf{v}}} + \beta {{\bf{d}}_{\cal S}}\left( {{{\bf{v}}}} \right)} \right)} \right)
\end{align}
where the retraction operation has been included to satisfy the constraint ${\cal S}$ and $\beta$ is a constant step length. To ensure monotonic updates, the selection of $\alpha$ and $\beta$ should satisfy the following principle \cite{alhujaili2019transmit}:
\begin{align}
\alpha \ge \frac{K}{8}{\lambda _{\max }}\left( {\bf{U}} \right),~~0 < \beta  < \frac{1}{{{\lambda _{\max }}\left( {\bf{U}} \right) + \alpha }},
\end{align}
where $\lambda _{\max }(\cdot)$ denotes the largest eigenvalue of its argument. By respectively setting $\bf{v}$, $\bf{U}$, $\rho$, and $K$ to the corresponding variables in (\ref{eq:w_optimize}) and (\ref{eq:theta_optimize}), $\mb{w}_{t+1}$ and $\bm{\theta}_{t+1}$ can be alternatingly optimized until the convergence of $\Delta I_{t+1}$. Consequently, we obtain the whole process of ARMO scheme in {\bf Algorithm 1}.

Thanks to the smooth objective function and gradient as well as the monotonic increase in each step, this iterative optimization converges on a compact manifold, driving the solution to a stationary point. As the computational complexity of ARMO mainly depends on gradient computation, the overall complexity of {\bf Algorithm 1} is ${\cal O}(Q{I_o}({I_w}{M^2} + {I_\theta }{N^2}))$, where ${I_w}$, ${I_\theta }$, and $I_o$ are the required number of iterations for the convergence of ${\bf w}_{t+1}$, ${\bm \theta}_{t+1}$, and $\Delta I_{t+1}$, respectively. This complexity is of the same order of magnitude as that of most RIS precoding algorithms.

\subsection{Acquisition of Channel Kernel}\label{subsec:Kernel_Acq}
The implementation of ARMO scheme requires the knowledge of the kernel ${\bm \Sigma}_{\bf h}$. In real-world 5G New Radio systems, to enable minimum mean squared error (MMSE) channel estimation, ${\bm \Sigma}_{\bf h}$ is usually acquired via channel state information reference signals (CSI-RSs) \cite{Sungwoo'18, Karthik'18}. To extend this process to RIS-aided systems, we introduce an adaptive kernel training strategy, which smoothly integrates kernel learning into channel estimations without additional resources. 

As a second-order statistic, the channel covariance ${\bm \Sigma}_{\bf h}$ varies much more slowly than the instantaneous channel $\bf h$. Consider a sliding window consisting of consecutive $R$ frames, where the inter-frame channels ${\bf h}_r$ for all $r\in\{1,\cdots,R\}$ satisfy distribution ${\cal CN}({\bf 0},{\bm \Sigma}_{\bf h})$. Our strategy is iteratively estimating the channels ${\bf h}_r$ and updating the sample kernel $\hat{\bm \Sigma}_{\bf h}$, such that the accuracy of estimator $\hat{\bf h}_r$ 
can be improved and $\hat{\bm \Sigma}_{\bf h}$ converges to the real kernel ${\bm \Sigma}_{\bf h}$ gradually. Let $\hat{\bm \Sigma}_{\bf h}^{(r)}$ denote the sample kernel in the $r$-th frame. Such a working flow process can be described as:
\begin{align}\label{eq:flow}
\notag
		\hat{\mb \Sigma }_{\bf h}^{(0)} &\underbrace{\overset{\rm ARMO}{\rightarrow} \hat{\mb{h}}_{1} \overset{\eqref{eq:kernel_learning}}{\rightarrow} 
			\hat{\mb \Sigma }_{\bf h}^{(1)}}_{\rm frame\:1} 
		\underbrace{\overset{\rm ARMO}{\rightarrow} \hat{\mb{h}}_{2} \overset{\eqref{eq:kernel_learning}}
        {\rightarrow} 
			\hat{\mb \Sigma }_{\bf h}^{(2)} }_{\rm frame\:2} \overset{\rm ARMO} 
             {\rightarrow} \\
		& \cdots 
		\underbrace{\overset{\rm ARMO}{\rightarrow} 
			\hat{\mb{h}}_{R} \overset{\eqref{eq:kernel_learning}}{\rightarrow}  \hat{\mb \Sigma }_{\bf h}^{(R)} }_{{\rm frame}\:R}\overset{\rm ARMO}{\rightarrow} \cdots,
\end{align}
where the sample kernel $\hat{\bm \Sigma}_{\bf h}^{(r)}$ is updated according to the estimated channel $\hat{\bf h}_r$. The following $R$-length sliding window is utilized to update the sample kernel in real time:
\begin{equation}\label{eq:kernel_learning}
\hat{\bm \Sigma}_{\bf h}^{(r)} = \begin{cases}
    \frac{{r - 1}}{r}\hat{\bf{\Sigma }}_{\bf{h}}^{(r - 1)} + \frac{1}{r}{\hat{\bf{h}}_r}\hat{\bf{h}}_r^H, & \text{if } r \le R, \\\frac{1}{R}
    \sum\nolimits_{r' = r - R}^r {{{\hat {\bf{h}}}_{r'}}\hat {\bf{h}}_{r'}^H}, & \text{if } r > R,
\end{cases}
\end{equation}
where the case when $r>R$ aims to track the time-varying ${\bm \Sigma}_{\bf h}^{(r)}$ in future frames. To trigger the working flow (\ref{eq:flow}), we set the initial
kernel as an identity matrix, i.e., $\hat{\mb \Sigma }_{\bf h}^{(0)}={\bf I}_{MN}$.

The proposed adaptive kernel learning strategy offers two main advantages. First, by initializing kernels as identity matrices, the method eliminates the need for prior knowledge of the actual kernels, rendering it suitable for a wide range of practical communication systems. Second, it enables simultaneous channel estimation and kernel training within a single frame by utilizing channels estimated by the ARMO algorithm for kernel training. This integration removes the need for a separate time period dedicated to kernel learning, significantly simplifying the frame structure and protocol in practice.

\section{Simulation Results}\label{sec:sim}
%In this section, simulation results are carried out to verify the effectiveness of the proposed ARMO scheme. 
%\subsection{Simulation Setup and Baselines}
We consider the uplink channel estimation of an RIS-aided DAS. Without loss of generality, the uniform linear array (ULA) is equipped on the \ac{bs} and the RIS is a uniform planar array (UPA). The dipole
antennas are considered for the \ac{bs} and the user, and patch antennas are deployed at the RIS. The electromagnetic mutual coupling effects among antennas/elements are calculated using Matlab Antenna Toolbox \cite{MathWorks5GToolbox}, and the clustered delay line (CDL)-A channel model defined in 3GPP TR 38.901 is adopted for simulations \cite{cdl}. Otherwise specifically
specified, the number of \ac{bs} antennas and that of RIS elements are set as $M=4$ and $N=8\times 8$, respectively. The antenna/element spacing is set to be ${\lambda}/{4}$. The \ac{snr} is defined as ${\rm SNR} = 1/{\sigma^2}$, whose default value is set to 10 dB. The evaluation criterion of estimation accuracy is the \ac{nmse}, which is expressed as ${\rm NMSE} = {\mathsf E}({\|{\bf h}-{\hat{\bf h}}\|^2}/{\|{\bf h}\|^2})$. The default value of pilot length is set to $Q=200$.

\begin{figure}[!t]
	\centering
	\includegraphics[width=0.5\textwidth]{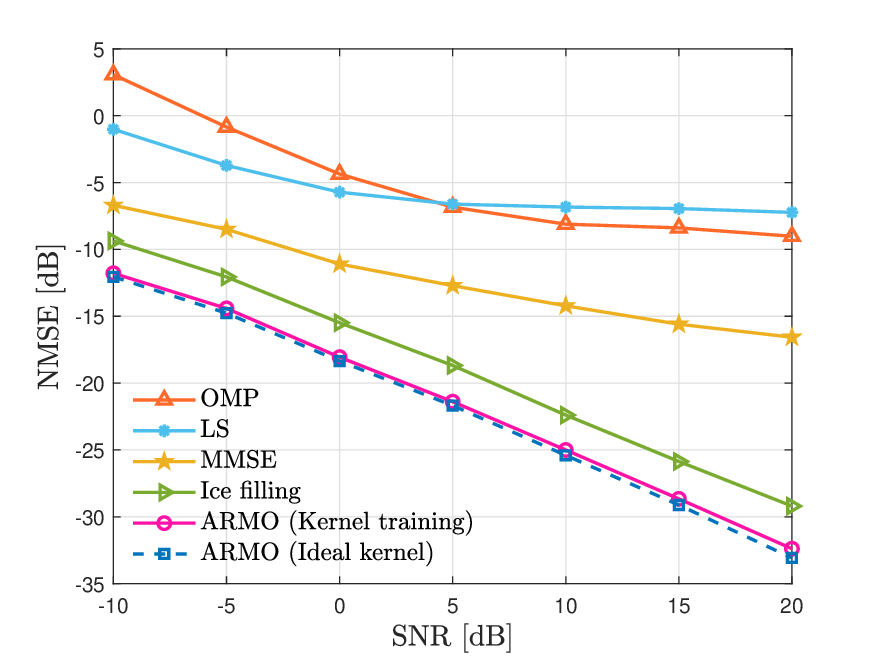}
    \vspace{-1em}
	\caption{NMSE performance versus SNR for different schemes.
	}
	\label{img:sim_NMSE_SNR}
\end{figure}

To verify the effectiveness of the proposed ARMO based channel estimator, the following schemes are compared:
\begin{itemize}
	\item {\bf LS}: The classical least-square (LS) scheme is used for RIS channel estimation, i.e., $\hat{\bf h}_{\rm LS}=({\bf X}_{\rm LS}^H)^{\dagger}{\bf y}_{\rm LS}$. The \ac{bs} combiner and the RIS phase-shift matrix in the observation matrix ${\bf X}_{\rm LS}$ are generated from DFT vectors.
	\item {\bf MMSE}: Assume the kernel ${\bm \Sigma}_{\bf h}$ is perfectly known. Using the same setting as the LS estimator, the MMSE estimator in \cite{Nadeem'20} is adopted to recover RIS channel $\bf 
	h$ via (\ref{eq:postmean}).
	\item {\bf OMP}: The orthogonal matching pursuit (OMP) method proposed in \cite{wei2021channel} is used to recover RIS channel $\bf h$. To satisfy the restricted isometry property, the \ac{bs} combiner and the RIS phase-shift matrix are randomly generated.
	\item {\bf Ice filling}: Assume the kernel ${\bm \Sigma}_{\bf h}$ is perfectly known and the RIS phase-shift matrix is randomly generated. The ice filling scheme in 
	\cite{cui2024nearoptimal} is used to design the \ac{bs} combiner. Then, $\bf 
	h$ is estimated via the MMSE estimator (\ref{eq:postmean}).
	\item {\bf ARMO (Ideal kernel)}: Assume ${\bm \Sigma}_{\bf h}$ is perfectly known. The proposed ARMO scheme in {\bf Algorithm \ref{alg:ARMO}} is employed to jointly design ${\bf X}$. Based on ${\bf X}$ and ${\bm \Sigma}_{\bf h}$, the MMSE estimator in  (\ref{eq:postmean}) is employed to recover $\bf h$.
	\item {\bf ARMO (Kernel training)}: The working flow process in (\ref{eq:flow}) and (\ref{eq:kernel_learning}) is employed to acquire and update the kernel $\hat{\bm \Sigma}_{\bf h}$. For kernel tracking, the length of the sliding window is set to $R=100$. Then, $\hat{\bm \Sigma}_{\bf h}$ is used as the input of {\bf Algorithm \ref{alg:ARMO}}. Based on the designed $\hat {\bf X}$ and $\hat{\bm \Sigma}_{\bf h}$, the MMSE estimator in  (\ref{eq:postmean}) is employed to recover $\bf h$.
\end{itemize}

%\subsection{Estimation Accuracy versus SNR and Pilot Length}

Firstly, we plot the NMSE as a function of SNR in Fig.~\ref{img:sim_NMSE_SNR}. It can be observed that, compared with the estimators without utilizing the prior information of kernels (LS and OMP), the schemes utilizing kernels (MMSE, ice filling, and ARMO) demonstrate their advantage in estimation accuracy. For example, thanks to the carefully designed combiners at the \ac{bs}, the ice filling scheme outperforms the MMSE scheme whose \ac{bs} combiners are generated from the DFT matrix. Furthermore, since our proposed ARMO schemes jointly design the \ac{bs} combiners and the phase-shift matrix at the RIS, they can achieve higher performance gains than the ice filling scheme that only focuses on the \ac{bs} combiner design. Besides, it is worth noting that, the ARMO scheme with kernel training achieves a very similar performance to that with perfect kernel knowledge. It indicates that, by alternatingly updating the kernel and channels, the knowledge of the real kernel is no longer necessary for the implementation of ARMO in practice.

\begin{figure}[!t]
	\centering
	\includegraphics[width=0.5\textwidth]{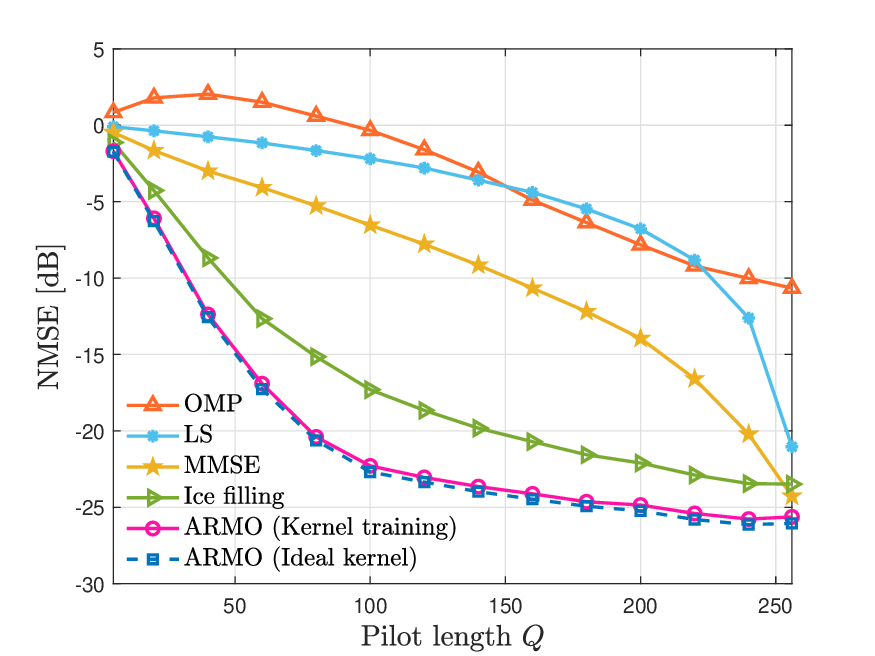}
	\vspace*{-1em}
	\caption{NMSE performance versus pilot length $Q$ for different schemes.
	}
    %\vspace*{-1em}
	\label{img:sim_NMSE_Q}
\end{figure}

Then, we plot the NMSE as a function of pilot length $Q$ in Fig. \ref{img:sim_NMSE_Q}. One can find that, with the assistance of the kernel, the kernel-related schemes can achieve considerable performance with a small number of pilots. In particular, the proposed ARMO schemes have an obvious superiority in estimation accuracy. For example, to achieve an NMSE of -10 dB, the required pilot lengths $Q$ for OMP, LS, MMSE, ice filling, ARMO (kernel training), and ARMO (ideal kernel) are 240, 225, 150, 45, 30, and 30, respectively. Note that, the OMP estimator does not perform well. It is because the small antenna spacing leads to the spatial oversampling and mutual coupling among antennas, thus the channels of DASs no longer enjoy the angular-domain sparse property. Furthermore, when the pilot length is sufficiently large (e.g., $Q=256$), the kernel-related schemes converge to a similar estimation accuracy. This occurs because the observation matrix of the MMSE estimator becomes a complete DFT matrix, which can fully characterize all channel patterns of DASs. Under these conditions, the accuracy gain from a carefully designed observation matrix becomes marginal.

\section{Conclusions}\label{sec:con}
This paper proposed an observation matrix design framework to enhance channel estimation in RIS-aided DASs. By formulating the design as an \ac{mi} maximization problem within a Bayesian framework, we jointly optimized the receiver combiners and RIS phase-shift matrices using ARMO scheme. In addition, an adaptive kernel learning strategy was introduced to enable simultaneous channel estimation and kernel training. Simulation results verified that the proposed ARMO-based channel estimator significantly outperforms existing schemes in estimation accuracy and pilot efficiency.

% \appendices

% \section{Proof of Lemma \ref{lemma:Sigma_decomp}}\label{appendix:Sigma_decomp}

\footnotesize
%\balance 
\bibliographystyle{IEEEtran}
	
%\begin{spacing}{1.8}
%\bibliography{IEEEabrv,reference}	
%\end{spacing}
	
\bibliography{IEEEabrv,reference}

\end{document}